\documentclass[journal,compsoc]{IEEEtran}
\usepackage{amsmath,amsfonts}
\usepackage{algorithmic}
\usepackage{array}
\usepackage{textcomp}
\usepackage{stfloats}
\usepackage{url}
\usepackage{verbatim}
\usepackage{graphicx}
\usepackage{balance}
\usepackage{tabularx}
\usepackage{multirow}
\usepackage{xcolor}
\usepackage{pifont}

\newcommand{\cmark}{{\ding{51}}}
\newcommand{\cmarkbold}{{\ding{52}}}
\newcommand{\xmark}{{\ding{55}}}
\newcommand\tablearraystrech{1.3}

\begin{document}

% \title{\lowercase{emo}DARTS: Enhancing Speech Emotion Recognition Through\\ Differentiable Architecture Search}
\title{\lowercase{emo}DARTS: Joint Optimisation of CNN \&
Sequential Neural Network Architectures for Superior Speech Emotion Recognition}
\author{Thejan~Rajapakshe$^{*}$,~\IEEEmembership{Student Member, IEEE},
        Rajib~Rana,~\IEEEmembership{Member, IEEE}, \\
        Sara~Khalifa,~\IEEEmembership{Member, IEEE},
        Berrak~Sisman,~\IEEEmembership{Member, IEEE},\\
        Bj\"{o}rn W.\ Schuller,~\IEEEmembership{Fellow, IEEE},
        and~Carlos Busso,~\IEEEmembership{Fellow, IEEE}\\% <-this % stops a space

\thanks{T.\ Rajapakshe and R.\ Rana are affiliated with the University of Southern Queensland, Australia.}
\thanks{S.\ Khalifa is affiliated with Queensland University of Technology, Australia.}
\thanks{B.\ Sisman is affiliated with The University of Texas at Dallas, USA.}
\thanks{B.\ Schuller is with CHI -- Chair of Health Informatics, MRI, Technical University of Munich, Germany, Munich Data Science Institute (MDSI), Germany, Munich Center for Machine Learning (MCML), Germany and GLAM -- Group on Language, Audio, \& Music, Imperial College London, UK}
\thanks{C.\ Busso is affiliated with The University of Texas at Dallas, USA.}%
\thanks{\protect $^{*}$Thejan.Rajapakshe@unisq.edu.au}% <-this % stops an unwanted space
% \thanks{Manuscript received January 25, 2024; revised August 16, 2021.}%
\thanks{Manuscript submitted to IEEE Transactions on Affective Computing on February 19, 2024}%
}

\maketitle

\begin{abstract}
% Speech Emotion Recognition (SER) is critical in enabling emotion-aware communication in human-computer interactions. Recent Deep Learning (DL) developments have significantly improved the performance of SER models by increasing model complexity. However, creating an optimum DL architecture requires prior expertise and experimental assessments. Neural Architecture Search (NAS), on the other hand, provides a potential path for automatically determining an ideal DL model. Differentiable  Architecture Search (DARTS), in particular, is an efficient technique to discover optimal models. This research introduces emoDARTS, a DARTS-optimised joint CNN and Sequential Neural Network (SeqNN: LSTM, RNN) architecture that enhances SER performance\CB{. The literature informs the selection of CNN and LSTM coupling to deliver improved performance.}
% % , where the literature informs the selection of CNN and LSTM coupling to deliver improved performance. 
% While DARTS has previously been used to choose CNN and LSTM operations independently, our technique adds a novel mechanism in selecting CNN and SeqNN operations in conjunction using DARTS. Unlike earlier work, we do not impose limits on the layer order of the CNN. Instead, we let DARTS choose the best layer order inside the DARTS cell on its own. We show that emoDARTS outperforms humans in designing CNN-LSTM models and surpasses the best-reported SER results achieved through DARTS on CNN-LSTM by evaluating our approach on the IEMOCAP, MSP-IMPROV, and MSP-Podcast datasets.

Speech Emotion Recognition (SER) is crucial for enabling computers to understand the emotions conveyed in human communication. With recent advancements in Deep Learning (DL), the performance of SER models has significantly improved. However, designing an optimal DL architecture requires specialised knowledge and experimental assessments. Fortunately, Neural Architecture Search (NAS) provides a potential solution for automatically determining the best DL model. The Differentiable Architecture Search (DARTS) is a particularly efficient method for discovering optimal models. This study presents emoDARTS, a DARTS-optimised joint CNN and Sequential Neural Network (SeqNN: LSTM, RNN) architecture that enhances SER performance. The literature supports the selection of CNN and LSTM coupling to improve performance. 

While DARTS has previously been used to choose CNN and LSTM operations independently, our technique adds a novel mechanism for selecting CNN and SeqNN operations in conjunction using DARTS. Unlike earlier work, we do not impose limits on the layer order of the CNN. Instead, we let DARTS choose the best layer order inside the DARTS cell. We demonstrate that emoDARTS outperforms conventionally designed CNN-LSTM models and surpasses the best-reported SER results achieved through DARTS on CNN-LSTM by evaluating our approach on the IEMOCAP, MSP-IMPROV, and MSP-Podcast datasets.

\end{abstract}


\section{Introduction}

\IEEEPARstart{R}{ecognising} the emotional nuances embedded in speech is a fundamental, yet complex challenge. Over the last decade, the field of Speech Emotion Recognition (SER) has experienced significant strides, predominantly driven by the exponential growth of deep learning~\cite{Zhao2019SpeechNetworks, Jalal2020EmpiricalRecognition, Lieskovska2021AMechanism, Latif2022MultitaskRecognition}. A key breakthrough facilitated by deep learning is its capability to automatically learn features, departing from the traditional reliance on manually crafted features shaped by human perceptions of speech signals. Nevertheless, determining the optimal deep-learning architecture for SER remains a challenging task that warrants attention. Conventional approaches involve iterative modifications and recursive training of models until an optimal configuration is found. However, this approach becomes prohibitively time-consuming due to the extensive training and testing required for numerous configurations.

An alternative to the conventional approach is the ``Neural Architecture Search'' (NAS), which can help discover optimal neural networks for a given task. The idea is to find the models' architecture to minimise the loss. In NAS, search is done over a discrete set of candidate operations, which requires the model to be trained on a specific configuration before moving on to the next configuration. Nevertheless, this approach demands considerable time and resources.
% This is, however, time-consuming and resource hungry. 
% The differentiable architecture search (DARTS)~\cite{Liu2018DARTS:SEARCH} found a way of relaxing the discrete set of candidate operations, allowing the space to be continuous. DARTS is able to decrease the computation time from 2\,000 GPU days of reinforcement learning or 3\,150 GPU days of evolution algorithm to 2–3 GPU days~\cite{Liu2018DARTS:SEARCH, Zoph2017LearningRecognition, Real2018RegularizedSearch}. Furthermore, through model optimisation, DARTS is expected to offer significantly high SER accuracy, which is cuurently quite low and needs improvement. These two points serve as motivation to use DARTS in this paper. 
%This paper uses DARTS for SER.

The Differentiable Architecture Search (DARTS) is a method that has been developed to optimise the search for a neural network architecture. It allows for the relaxation of the discrete set of candidate operations, making the space continuous and reducing the computation time significantly, from 2,000 GPU days to just 2-3 GPU days. This is a major improvement from the previous methods of reinforcement learning or evolution algorithm, which required 2,000 and 3,150 GPU days, respectively. Additionally, through network optimisation DARTS has the potential to offer significantly high SER accuracy, which is currently quite low and needs improvement. These two points serve as motivation to use DARTS for SER.

\begin{figure}
    \centering
    \includegraphics[width=.90\linewidth]{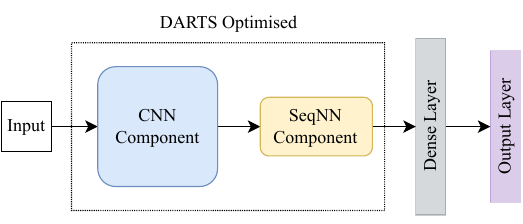}
    \caption{The proposed architecture of emoDARTS passes the input features to the CNN component through the SeqNN component and finally to a dense layer. The optimum CNN and SeqNN operations are selected by DARTS jointly.}
    \label{fig:overall_architecture_intro}
\end{figure}

% It was previously demonstrated that 
Additionally, previous studies have shown that a multi-temporal Convolutional Neural Network (CNN) stacked on a Long Short-Term Memory Network (LSTM)
%BS: I think, LSTM has not yet been introduced ias abbreviation - not even in the abstract
can capture contextual information at multiple temporal resolutions, complementing LSTM for modelling long-term contextual information, thus offering improved performance~\cite{Han2018TowardsRecognition, Haque2018ImageAttention, Li2019ImprovedLearning, Latif2022SelfRecognition}. Sequential Neural Networks (SeqNN) like Recurrent Neural Networks (RNN)
%BS: RNN has also not been introduced as abbreviation - please decide if you feel these are too common to introduce and otherwise consequently introduce abbreviations.
or LSTM can easily identify the patterns of a sequential stream of data. 
This paper takes a pioneering step by leveraging DARTS for a novel joint CNN--SeqNN configuration, named ``emoDARTS'', as depicted in Figure~\ref{fig:overall_architecture_intro}, with an attention network seamlessly integrated into the SeqNN component to further elevate its performance. 

%\textcolor{blue}{SER needs a performance boost as motivation }

The investigation of DARTS within the SER domain is minimal and invites further inquiry to uncover the potential for improving SER performance. 
% Only recently -- since last year~\cite{Sun2023EmotionNAS:Recognition, Wu2022NEURALRECOGNITION}, researchers have used DARTS in SER tasks to optimise their models. 
DARTS has only recently been employed in SER tasks to improve models, as recently as 2022~\cite{Sun2023EmotionNAS:Recognition, Wu2022NEURALRECOGNITION}, wherein the researchers have mostly applied DARTS separately on CNNs and RNNs~\cite{Xu2020PC-DARTS:Search, Biju2019ElectricDARTS, Maulik2020RecurrentEmulation}. Hence, the viability of utilising DARTS jointly for CNN and SeqNN requires exploration. While there is a lone study that explores the joint optimisation of CNN and LSTM~\cite{Wu2022NEURALRECOGNITION}, it imposes constraints on the layer order for the CNN within the DARTS component, thereby limiting the full potential of DARTS. In response to this limitation, our paper takes on the challenge of optimising this joint configuration without such constraints. The contributions of this paper are summarised as follows.

\begin{enumerate}
    \item This paper proposes a novel DARTS-optimised joint CNN and SeqNN architecture, emoDARTS, achieving greater autonomy to DARTS in selecting optimal network configurations.
    % novel method for neural network optimisation using DARTS for SER tasks.
    % \item While most existing literature employing DARTS in SER has predominantly focused on IEMOCAP, our study extends this scope by incorporating the MSP-IMPROV and MSP-Podcast datasets. This broader dataset selection showcases the robust generalisation capabilities of the proposed emoDARTS model.  
    \item We demonstrate the robust generalisation capabilities of the proposed emoDARTS model by testing it on three widely used datasets: IEMOCAP, MSP-IMPROV, and MSP-Podcast.
    % \textcolor{blue}{this can stay here, but it doesn't look like a strong novelty (the use of 2 more datasets). Is there any other point that we can include in this section?}
    \item  Experimental results demonstrate that emoDARTS
achieves 
%BS: use "significantly" if you actually provide a significance testing method and p-value, please!
%TR significantly
considerably higher SER accuracy than humans designing the CNN-LSTM configuration. It also outperforms the best-reported SER results achieved using DARTS on CNN-LSTM.
\end{enumerate}

\section{Related Work}
 % \textcolor{blue}{number of references is relatively low in this paper (for a TAC submission), can you make sure we cite all the related work?}
This section delves into the existing literature on using DARTS and NAS for SER. Notably, our exploration reveals a limited number of papers in this space. We therefore extend our review to encompass relevant papers in related fields to provide a comprehensive perspective. For completeness, we also include studies employing CNNs, LSTM networks, and their joint utilisation for SER.

\subsection{Speech Emotion Recognition using CNN and LSTM} 
One of the earliest uses of CNN networks in SER is reported by Zheng et al.\ in 2015~\cite{Zheng2015AnNetworks}.
The authors introduced a spectrum generated from an audio signal to a CNN network and output the recognised emotion. 
The authors report that they can surpass the SVM-based classification performance and reach 40\% classification accuracy for a five-class classification using the IEMOCAP dataset.

The earliest work combining CNN and LSTM for SER is by Trigeorgis et al.\ in 2016~\cite{Trigeorgis16-AFE}. The authors show an impressive improvement by a fully self-learnt representation over traditional expert-crafted features on dimensional emotion recognition.

Zhaoa et al.~\cite{Zhao2019SpeechNetworks} show that using CNN and LSTM networks combined in the same SER model produces better results than using only CNN. Using the IEMOCAP dataset, they obtained a speaker-independent accuracy of 52\% by using a log-Mel spectrogram as the input feature. Their SER approach utilises an LSTM layer to learn contextual dependencies in local features, while a CNN-based layer learns the local features.

\subsection{Application of NAS and DARTS in SER and Related Fields} 

The first paper suggesting NAS in SER was by Zhang et al.\ in 2018~\cite{Zhang18-ELF}. The authors employ a controller network that shapes the architecture by the number of layers and nodes per layer and the hyperparameter activation function of a child network by reinforcement learning. They show an improvement over human-designed architectures and random searches of these.

% Real E. et al.~\cite{Real2019AgingSearch} use an evolutionary algorithm to develop an image classification model called `\textit{AmoebaNet}-A'. They introduce a new property, `age', to the tournament selection evolutionary algorithm to favour the younger genotype, which allowed them to achieve 83.9\% top-1 \textit{ImageNet} accuracy.

Zoph and Le~\cite{Zoph2016NeuralLearning} use reinforcement learning to optimise an RNN network that develops model architectures to maximise the resulting accuracy of the generated model. As a result, they develop outstanding models for the CIFAR-10 and Penn Treebank datasets. They were able to develop a convolutional network architecture for the CIFAR-10 dataset which has a $3.65$ error rate and a recurrent network architecture for Penn Treebank with $62.4$ perplexity.

Even though NAS is primarily used to find optimised architecture for complex and large models, researchers have also studied the possibility of using NAS to design smaller deep neural network models. Liberis et al.~\cite{Liberis2020NAS:Microcontrollers} develop a NAS system called $\mu$NAS to design smaller neural architectures that can run on microcontroller units. They improve the top-1 accuracy by $4.8\%$ in image classification problems while reducing the memory footprint up to 13 times. Similarly, Gong et al.~\cite{Gong2019MixedLearning} study the feasibility of using NAS for reducing deep learning models to deploy on resource-constrained edge devices. 

Traditional NAS consumes much computational power and time to achieve the optimal model for a given problem. In 2018, Liu et al.~\cite{Liu2018DARTS:SEARCH} came up with a differentiable approach to solving the optimisation by continuous relaxation of the architecture representation. This approach is more compute efficient and high performing as the search space is not discrete and non-differentiable. They produce high-performing CNN and RNN architectures for tasks such as image recognition and language modelling within a fraction of the search cost of traditional NAS algorithms. DARTS has been popular in the past three years with many studies carried on for extending and improving the algorithm~\cite{Liu2018ProgressiveSearch,Liang2019DARTS+:Stopping,Xu2020PC-DARTS:Search,Wan2020FBNetV2:Dimensions,Chen2020StabilizingRegularization,Yu2023CyclicSearch}.

Wu et al.~\cite{Wu2022NEURALRECOGNITION} proposed a uniform path dropout strategy to optimise candidate architecture. They use SER as their DARTS application and the IEMOCAP dataset to develop an SER model with an accuracy of $56.28\%$ for a four-class classification problem using discrete Fourier transform spectrograms extracted from audio as input. In their work, the authors specify layer order as two convolution layers at first, followed by a max-pooling layer, a convolution layer. They use DARTS to select the optimum parameters for each layer. We, on the other hand, do not specify the layer sequence and instead enable DARTS to select the ideal design with minimal interference.

% TABLE START
\begin{table*}[!t]
\centering
\caption{Summary and focus on the literature on NAS, DARTS, and speech emotion recognition.}
\label{tab:literature}
\bgroup
\def\arraystretch{\tablearraystrech}
\begin{tabular}{|l|cccc|ccc|}
\hline
\multirow{2}{*}{Paper}                                 & \multicolumn{4}{c|}{Focus}   & \multicolumn{3}{c|}{Dataset}                                                                              \\ \cline{2-8} 
                                                       & \multicolumn{1}{c|}{SER} & \multicolumn{1}{c|}{NAS} & \multicolumn{1}{c|}{DARTS} & \multicolumn{1}{c|}{Joint Opt. of CNN \& SeqNN.} & \multicolumn{1}{c|}{IEMOCAP}& \multicolumn{1}{c|}{MSP-IMPROV}& \multicolumn{1}{c|}{MSP-Podcast} \\ \hline

Zoph and Le 2016 \cite{Zoph2016NeuralLearning}          & \multicolumn{1}{c|}{\xmark}                          & \multicolumn{1}{c|}{\cmark}                          & \multicolumn{1}{c|}{\xmark}                          & \multicolumn{1}{c|}{\xmark}   & \multicolumn{1}{c|}{\xmark}   & \multicolumn{1}{c|}{\xmark}   & \multicolumn{1}{c|}{\xmark}        \\ \hline
Zhang et al. 2018 \cite{Zheng2015AnNetworks}              & \multicolumn{1}{c|}{\cmark}                          & \multicolumn{1}{c|}{\cmark}                          & \multicolumn{1}{c|}{\xmark}                          & \multicolumn{1}{c|}{\xmark}   & \multicolumn{1}{c|}{\cmark}   & \multicolumn{1}{c|}{\xmark}   & \multicolumn{1}{c|}{\xmark}        \\ \hline
Liu et al. 2018 \cite{Liu2018DARTS:SEARCH}                  & \multicolumn{1}{c|}{\xmark}                          & \multicolumn{1}{c|}{\cmark}                          & \multicolumn{1}{c|}{\cmark}                          & \multicolumn{1}{c|}{\xmark}   & \multicolumn{1}{c|}{\xmark}   & \multicolumn{1}{c|}{\xmark}   & \multicolumn{1}{c|}{\xmark}        \\ \hline
Gong et al. 2019 \cite{Gong2019MixedLearning}              & \multicolumn{1}{c|}{\xmark}                          & \multicolumn{1}{c|}{\cmark}                          & \multicolumn{1}{c|}{\xmark}                          & \multicolumn{1}{c|}{\xmark}   & \multicolumn{1}{c|}{\xmark}   & \multicolumn{1}{c|}{\xmark}   & \multicolumn{1}{c|}{\xmark}        \\ \hline
Liberis et al. 2020 \cite{Liberis2020NAS:Microcontrollers} & \multicolumn{1}{c|}{\xmark}                          & \multicolumn{1}{c|}{\cmark}                          & \multicolumn{1}{c|}{\xmark}                          & \multicolumn{1}{c|}{\xmark}   & \multicolumn{1}{c|}{\xmark}   & \multicolumn{1}{c|}{\xmark}   & \multicolumn{1}{c|}{\xmark}        \\ \hline
Wu et al. 2022 \cite{Wu2022NEURALRECOGNITION}              & \multicolumn{1}{c|}{\cmark}                          & \multicolumn{1}{c|}{\cmark}                          & \multicolumn{1}{c|}{\cmark}                          & \multicolumn{1}{c|}{\cmark}   & \multicolumn{1}{c|}{\cmark}   & \multicolumn{1}{c|}{\xmark}   & \multicolumn{1}{c|}{\xmark}        \\ \hline
Sun et al. 2023 \cite{Sun2023EmotionNAS:Recognition}     & \multicolumn{1}{c|}{\cmark}                          & \multicolumn{1}{c|}{\cmark}                          & \multicolumn{1}{c|}{\cmark}                          & \multicolumn{1}{c|}{\xmark}   & \multicolumn{1}{c|}{\cmark}   & \multicolumn{1}{c|}{\xmark}   & \multicolumn{1}{c|}{\xmark}        \\ \hline
\textbf{emoDARTS}                                             & \multicolumn{1}{c|}{\cmarkbold}                          & \multicolumn{1}{c|}{\cmarkbold}                          & \multicolumn{1}{c|}{\cmarkbold}                          & \multicolumn{1}{c|}{\cmarkbold}   & \multicolumn{1}{c|}{\cmarkbold}   & \multicolumn{1}{c|}{\cmarkbold}   & \multicolumn{1}{c|}{\cmarkbold}        \\ \hline
\end{tabular}
\egroup
\end{table*}
% TABLE END

EmotionNAS is a two-branch NAS strategy introduced by Sun et al.~\cite{Sun2023EmotionNAS:Recognition} in 2023. The authors use DARTS to optimise their two models in two branches, the CNN model and RNN model, which use a spectrogram and a waveform as inputs, respectively. They obtained an unweighted accuracy of $72.1\%$ from the combined model for the IEMOCAP dataset. They also report the performance of $63.2\%$ in the spectrogram branch, which only uses a CNN component.
The main difference between our approach and the study by Sun et al.~\cite{Sun2023EmotionNAS:Recognition} is that we use a SeqNN component coupled in series with the CNN layer as in Figure~\ref{fig:overall_architecture_darts} while Sun et al.~\cite{Sun2023EmotionNAS:Recognition} use an RNN layer in parallel to the CNN layer in a different branch. 

We conducted preliminary research to determine the feasibility of utilising DARTS for SER in a CNN-LSTM architecture, where we only optimised the CNN network using DARTS~\cite{Rajapakshe2023EnhancingSearch}. 
% Using DARTS to optimise the CNN component in a CNN-LSTM model yielded encouraging results, we achieved 73.2\% accuracy for the IEMOCAP dataset. 
This paper extends the idea of using DARTS in SER but with more relaxation in the SeqNN component by jointly optimising the whole architecture. %BS: perfect - thanks for adding this :)

In recent years, the literature has highlighted the use of attention networks in SER, which has provided superior outcomes~\cite{Zhang2018AttentionRecognition,Chen20183-DRecognition}. We added an attention network component to the DARTS search scope to discover whether it improves performance.
Zou et. al.~\cite{Zou2022SPEECHINFORMATION} have introduced a concept called `co-attention' where many separate inputs from multimodal inputs are fused by co-attention. They used three sets of features MFCC, spectrogram, and Wav2Vec2 features from the IEMOCAP dataset and obtained 72.70\% accuracy. 
Liu et al.~\cite{Liu2023SpeechLearning} have utilised an attention-based bi-directional LSTM followed by a CNN layer for a SER problem. They have achieved a significant performance of $66.27\%$ for the IEMOCAP Dataset. Their idea of `CNN - LSTM attention' paved the foundation for our model architecture.

% \TR{Lit about: 1. Attention in RNN/LSTM for SER}

In Table~\ref{tab:literature}, we briefly compare the existing studies with emoDARTS. The comparison clearly shows that,

\begin{enumerate}
\item While some studies employ NAS for SER, the utilisation of DARTS in SER is notably limited.
\item Singularly, one study has explored the concept of jointly optimising CNN and SeqNN using DARTS for SER, in which the researchers specified the layer order. However, in our study, we let DARTS determine the optimal network from a relaxed search scope which enables it to select any operation in search space at the optimum layer.
\item Most existing studies primarily focus on the IEMOCAP dataset. In contrast, our study uniquely incorporates three widely recognised SER datasets: IEMOCAP, MSP-IMPROV, and MSP-Podcast to demonstrate the generalisation power of emoDARTS.
\end{enumerate} %BS: this sounds a bit like a repetition from above - perhaps cut it above?

\section{\lowercase{emo}DARTS Framework}

\begin{figure}[!t]
    \centering
    \includegraphics[width=1\linewidth]{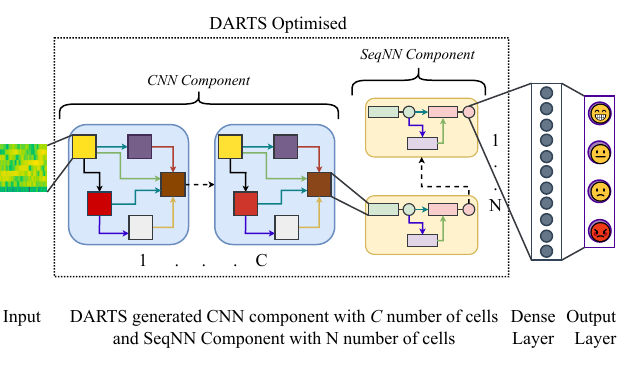} 
    \caption{The emoDARTS architecture comprises input features processed through CNN, SeqNN, and Dense layers and it utilises DARTS for jointly optimising the CNN and SeqNN components.}
    \label{fig:overall_architecture_darts}
\end{figure}

% emoDARTS applies DARTS to enhance SER using a CNN-SeqNN network, influenced by research demonstrating improved SER performance when combining CNN and LSTM layers~\cite{Han2018TowardsRecognition, Li2019ImprovedLearning, Latif2022SelfRecognition, Liu2023SpeechLearning}. 
% We model our network as a multi-component DARTS network where the input is fed into a CNN component and the output from the CNN component is fed into a SeqNN component, but both components are jointly optimised during the architecture search phase and an optimum architecture is obtained (Figure ~\ref{fig:overall_architecture_darts})
%BS: Added:
% As introduced above, 

The proposed `emoDARTS' uses DARTS to improve SER using a CNN-SeqNN network, which was motivated by studies that showed increased SER performance when CNN and LSTM layers were combined~\cite{Han2018TowardsRecognition, Li2019ImprovedLearning, Latif2022SelfRecognition, Liu2023SpeechLearning}. We represent our network as a multi-component DARTS network, with the input fed into a CNN component and the output from the CNN component fed into a SeqNN component, but all components are optimised jointly during the architecture search phase, delivering an optimal architecture (Figure~\ref{fig:overall_architecture_darts}).

\begin{figure*}[!t]
    \centering
    \includegraphics[width=0.85\textwidth]{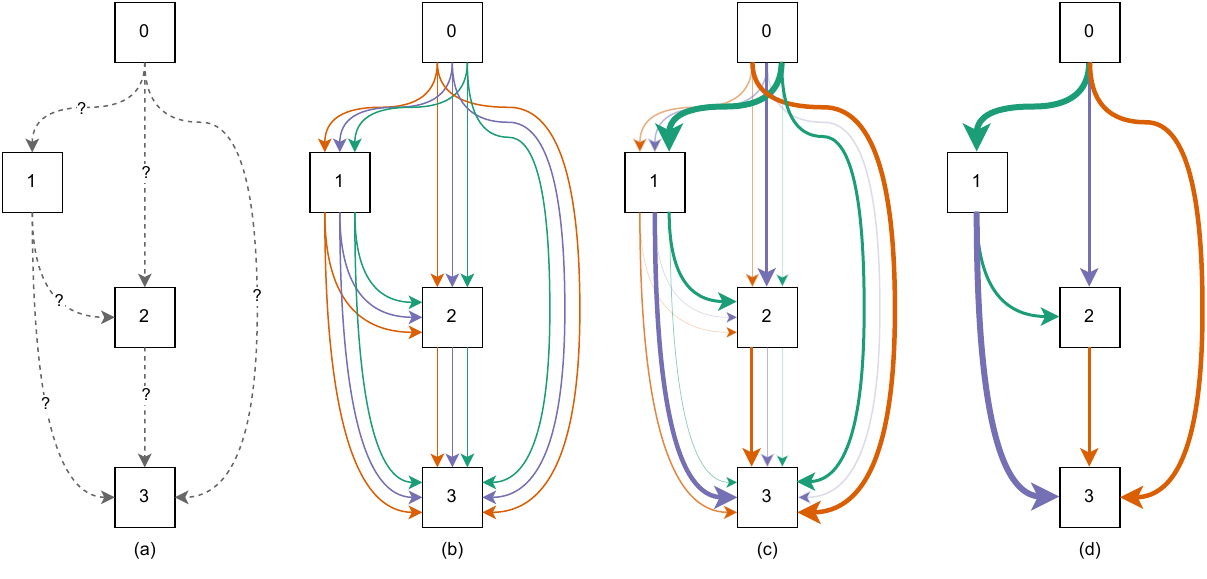}
    \caption{DARTS employs steps (a) to (d) to search cell architectures: (a) initialises the graph, (b) forms a search space, (c) updates edge weights, and (d) determines the final cell structure. Nodes signify representations, edges represent operations, with light-coloured edges indicating weaker and dark-coloured edges representing stronger operations.}
    \label{fig:DARTS_structure}
\end{figure*}

% DARTS uses a differentiable approach to network optimisation. The building block of the DARTS algorithm is a computation cell. It seeks to optimise the cell to gain maximum performance from the architecture. A DARTS cell is modelled as a directed graph where each node is a \TR{feature} representation, and \TR{an} edge is an operation that can be applied to a representation. 
DARTS uses a differentiable approach to network optimisation. A computation cell is the DARTS algorithm's fundamental unit. It aims to optimise the cell so that the architecture can function to its maximum performance. A DARTS cell is described as a directed graph, with each node representing a feature (representation) and each edge representing an operation that can be performed to a representation.
% To use the DARTS cell in a component, we model the node as a feature map and an edge for an operation. 
% One speciality of this graph is that each node connects by an edge with all its preceding nodes as in Figure
One unique feature of this network is that each node is connected to all of its previous nodes by an edge, as seen in Figure~\ref{fig:DARTS_structure} (a). If the output of the node $j$ is $x^{(j)}$ and the operation `$o$' on the edge connecting the nodes $i$ and $j$ is $o^{(i,j)}$, $x^{(j)}$ can be obtained by the Equation~\ref{eq:output_from_node}:
\begin{equation}
    \centering
    \label{eq:output_from_node}
    x^{(j)} = \sum_{i<j} o^{(i,j)}(x^{(i)})
\end{equation}

% Initially, the candidate search space is created by connecting each node of the DARTS cell with \TR{all the candidate operations (having multiple edges between nodes)} %a set of operations 
% as shown in Figure~\ref{fig:DARTS_structure} (b). A weight parameter $\alpha$ is introduced to Equation~\ref{eq:output_from_node} to find the optimum edge (operation) between two nodes, $i$ and $j$, out of the candidate search space of all the operations. The output from the node can be expressed as in Equation~\ref{eq:output_from_node_weighted}.
In the beginning, the candidate search space is generated by combining each node of the DARTS cell with all the candidate operations (with multiple links between nodes), as illustrated in Figure~\ref{fig:DARTS_structure} (b). Equation~\ref{eq:output_from_node} incorporates a weight parameter $\alpha$ to identify the optimal edge (operation) connecting two nodes, $i$ and $j$, from the candidate search space of all operations. Equation~\ref{eq:output_from_node_weighted} describes how the node's output be represented.
\begin{equation}
    \centering
    \label{eq:output_from_node_weighted}
    x^{(j)} = \sum_{i<j} \alpha^{(i,j)}o^{(i,j)}(x^{(i)})
\end{equation}
Then, the continuous relaxation of the search space updates the weights ($\alpha^{i,j}$) of the edges. The final architecture can be obtained by selecting the operation between two nodes with the highest weight $(o^{(i,j)*})$ by using Equation~\ref{eq:highest_weight_edge}.
\begin{equation}
    \centering
    \label{eq:highest_weight_edge}
    o^{(i,j)*} = argmax_o(\alpha^{(i,j)})
\end{equation}
The searched discrete cell architecture is shown in Figure~\ref{fig:DARTS_structure} (d).

% The number of cells ($C$ or $N$) in a component is a parameter to the DARTS algorithm which defines how many DARTS cells are stacked to create a component in the model. Each cell uses the output from the last two cells as the input. If output from each cell $t$ is $y_t$ and if the function inside the cell is $f$, $y_t$ can be expressed as; 
The number of cells ($C$ or $N$) in a component is a parameter for the DARTS algorithm that specifies how many DARTS cells are stacked to form a component in the model. Each cell takes the last two cells' output as input. If the output from each cell $t$ is $y_t$ and the function within the cell is $f$, then $y_t$ can be represented as;
\begin{equation}
    \centering
    \label{eq:layer_output}
    y_t = f(y_{t-1}, y_{t-2})
\end{equation}

% DARTS uses two types of CNN cells in the CNN component, namely `normal' and `reduction' cells. 
% It sets the stride as 1 in normal cells and 2 in reduction cells, so the output is down-sampled in reduction cells. This down-sampling enables the model to remove the redundancy of intermediate features and reduce the complexity.
DARTS' CNN component has two types of CNN cells: `normal' and `reduction' cells. It sets the stride to one in normal cells and two in reduction cells, resulting in a downsampled output in the reduction cells. This downsampling allows the model to eliminate the duplication of intermediate characteristics, reducing complexity.

% The reason we chose to jointly optimise the CNN and SeqNN components rather than separately is; that it is important for the components downstream (in this case CNN Component), to have knowledge about the behaviour of upstream components. Joint optimisation favours the architecture search in several ways such as 1. the back-propagation of loss minimisation goes through all the components in a single compute graph, 2. reduces the time required in the search phase as it searches the architecture of the whole network at once.

We decided to jointly optimise the CNN and SeqNN components rather than individually since it is important for downstream components (in this case, the CNN component) to understand the behaviour of upstream components (SeqNN). Joint optimisation in a multiple-component network improves architecture search in various ways, including: 1. the back-propagation of loss minimisation flows through all the components in a single compute graph; and 2. it reduces the time required in the search phase by searching the architecture of the whole network at once.

\section{Experimental Setup}

\subsection{Dataset and Feature Selection}
We use the widely used IEMOCAP~\cite{Busso2008IEMOCAP:Database}, MSP-IMPROV \cite{Busso2017MSP-IMPROV:Perception}, and MSP-Podcast~\cite{Lotfian2019BuildingRecordings} datasets for our experiments. 
Our study takes the improvised subset of IEMOCAP and the four categorical labels, happiness, sadness, anger, and neutral as classes from the datasets. %BS: "improvised" sounds not comparable to others, but I guess, it is - please clarify - also regarding the cross-validation.
We employ five-fold cross-validation with at least one speaker out in our training and evaluations. At each fold, the training dataset is divided into two subsets, `search', and `training', by a $70/30$ fraction. The `search' set is used in the architecture search; the `training' set is used in optimising the searched architecture, and the remaining testing dataset is used to infer and obtain the testing performance of the searched and optimised model. 
This way, we manage to split the dataset into three sets in each cross-validation session. %BS: unclear if strictly speaker independent! Please clarify.
The IEMOCAP dataset has five sessions with ten actors and two unique speakers in each. We use one session for the testing dataset and four sessions for the search and training datasets. Similarly, MSP-Improv comprises six sessions including twelve actors. We take one session in the testing dataset and the remaining five sessions in the search and training dataset. MSP-Podcast includes a speaker ID with each audio utterance, and we group the entire dataset by the speaker and divide it by the $70/30$ rule.

In this research, we use Mel Frequency Cepstral Coefficients (MFCC) as input features to the model. MFCC has been used as the input feature in many SER studies in the literature~\cite{Davis1980ComparisonSentences, Latif2019DirectSpeech} and has proven to obtain promising results. 
%BS: in DL, more often than MFCCs, MFBs are used - perhaps comment why MFCC and not MFB?
Some machine learning research uses the Mel Filter bank as an input feature when the algorithm is not vulnerable to strongly correlated data. We picked the MFCC for this study since the deep learning model is produced automatically and we do not want to infer the model's sensitivity to correlated input.
We extract $128$ MFCCs from each $8$-second audio utterance from the dataset. If the audio utterance length is less than $8$ seconds, we added padding with zeros while the lengthier utterances are truncated. The MFCC extraction from the Librosa python library ~\cite{McFee2015Librosa:Python} 
outputs a shape $128\times512$, downsampled with max pooling, to create a spectrogram of the shape $128\times128$.

\subsection{Baseline Models}
We compare the performance of emoDARTS for SER with three models developed without DARTS (w/o DARTS)
% human-designed (HD) %hand-engineered baseline models
: 1) CNN, 2) CNN+ LSTM, and 3) CNN+LSTM with attention
as baseline models. The CNN baseline model consists of a CNN layer (kernel size=2, stride=2, and padding=2) followed by a Max-Pooling layer (kernel size=2 and stride=2). Two dense layers then processes the output from the Max-Pooling layer after applying a dropout of $0.3$. Finally, the last dense layer has four output units resembling the four emotion classes, and the model outputs the probability estimation of each emotion for a given input by a Softmax function. 

\begin{figure}[ht]
    \centering
    \includegraphics[width=\linewidth]{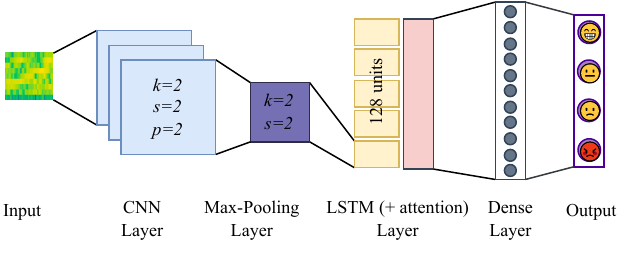}
    \caption{Visualisation of the CNN+LSTM attention baseline model. The parameters of the CNN layer are: kernel size (k)=2, stride (s)=2 and, padding (p)=2 and the parameters of the Max-pooling layer are: kernel size (k)=2 and stride (s)=2 and the LSTM layer has 128 units.}
    \label{fig:baseline_cnn_lstm_model}
\end{figure}

The CNN+LSTM baseline model is built, including an additional bi-directional LSTM layer of 128 units after the Max-Pooling layer. An attention layer is added to the LSTM layer in the `CNN+LSTM attention' baseline model. Figure~\ref{fig:baseline_cnn_lstm_model} shows the architecture of the CNN+LSTM attention baseline model

\subsection{DARTS Configuration}
We divide the cell search space operations into the two separate parts CNN and SeqNN based on the components they apply.
Table~\ref{tab:darts_operations} lists the type of operations used in each component. 
The cell search space of the CNN component consists of pooling operations such as $3\times3$ max pooling ($i=3$) and $3\times3$ average pooling ($i=3$), convolutional operations such as $3\times3$ and $5\times5$ separable convolutions ($i=3,5$), $3\times3$ and $5\times5$ dilated convolution ($i=3,5$), $7\times1 - 1\times7$ factorised convolution ($i=7$), identity connections, and no connections while the SeqNN component consists of operations such as RNN of layers 1 through 4 ($j=1,2,3,4$), RNN of layers 1 and 2 with attention ($j=1,2$), LSTM of layers 1 through 4 ($j=1,2,3,4$), LSTM of layers 1 and 2 with attention ($j=1,2$), identity connections and no connections. 

\renewcommand{\arraystretch}{\tablearraystrech}
\begin{table}
\caption{Type of DARTS operations used in each component. ``$i$'' represents the kernel size in CNN while ``$j$''  represents the number of layers in the SeqNN component.}
    \label{tab:darts_operations}
    \centering
    \begin{tabular}{l|l|p{0.45\linewidth}}
        \textbf{Component} & \textbf{Operation} & \textbf{Description} \\ \hline
        CNN & max\_pool\_$i$x$i$ &  Max Pooling layer with kernel $i$ \\ \hline
        CNN & avg\_pool\_$i$x$i$ &  Average Pooling layer with kernel $i$ \\ \hline
        CNN & dil\_conv\_$i$x$i$ &  Dilated Convolution layer with kernel $i$, and dilation 2 \\ \hline
        CNN & sep\_conv\_$i$x$i$ &  Two Convolution layers with kernel $i$ \\ \hline
        CNN & conv\_$i$x1\_1x$i$ &  Two Convolution layers with first kernel ($i$x1) and second (1x$i$) \\ \hline
        SeqNN & lstm\_$j$ & LSTM with $j$ layers \\ \hline
        SeqNN & lstm\_att\_$j$ & LSTM of $j$ layers with Attention \\ \hline
        SeqNN & rnn\_$j$ & RNN with $j$ layers \\ \hline
        SeqNN & rnn\_att\_$j$ & RNN of $j$ layers with Attention \\ 
    \end{tabular}    
\end{table}
\renewcommand{\arraystretch}{1}

We use stochastic gradient descent with a learning rate from $0.025$ to $0.001$
%BS: does the learning rate miss here? Please add.
using a cosine annealing schedule as the optimiser to optimise the weights of the operations. The search is run for $300$ epochs. 

In our experiments, we use four DARTS cells ($C=4$) for the CNN component following the work of Liu et al.~\cite{Liu2018DARTS:SEARCH} and two DARTS cells ($N=2$) for the SeqNN component. The intuition of using $N=2$ for the SeqNN component is discussed in section~\ref{sec:converging_to_local}.
As defined in~\cite{Liu2018DARTS:SEARCH}, we apply reduction cells at every \(\frac{1}{3}C^{th}\) and \(\frac{2}{3}C^{th}\) position of the layers in CNN component. We randomly initialise $\alpha$ values and the DARTS search algorithm optimises $\alpha$ values related to each operation.
The output from the CNN component is flattened to a vector before passing to the SeqNN component to adjust the input dimension of the RNN and LSTM layers. 

Once the search operation completes, it outputs the architecture of DARTS cells, which is called ``genome''. We create a deep learning model with the CNN component having four CNN cells and the SeqNN component having two SeqNN cells. This model is trained for 300 epochs with the training set of the datasets to minimise the loss.

% \subsection{emoDARTS Configuration}
% In emoDARTS, the output from the CNN component is passed through to the LSTM layer with 256 units as a vector after flatterning the 2D matrix from CNN.
% Attention is introduced into the LSTM layer by combining the attention module with the output of the LSTM layer.

We use the popular deep learning library PyTorch~\cite{Paszke2019PyTorch:Library} for model development and training. The experiments are run on an NVIDIA A100 GPU with 40\,GB of VRAM. We published the source code related to our research in a dedicated GitHub repository, allowing for smooth replication of our research findings\footnote{\url{https://github.com/iot-health/emoDARTS}}.

% \HL{compare the 3 studies with parameters, results etc. show the difference in the parameters, focus about layers, }

\section{Evaluation}
We report the results using the Unweighted Accuracy (UA\%), calculated by dividing the total of all classes' recall by their number. This is recognised to depict unbalanced data workloads intrinsic to SER accurately. We additionally provide the Weighted Accuracy (WA\%) mainly to compare our results with relevant studies~\cite{Sun2023EmotionNAS:Recognition,Wu2022NEURALRECOGNITION}. Last but not least, we also report the number of parameters of the model as an indication of the model's complexity, calculated by adding all trainable parameters in the created model.

\subsection{CNN only model}
We initially assess the performance of the CNN-only model generated by DARTS (CNN -- DARTS) compared to our benchmark model, specifically CNN -- w/o DARTS, using the IEMOCAP dataset. The results, detailed in Table~\ref{tab:baseline_vs_nas_cnn_only}, reveal that the DARTS-generated CNN model outperforms the performance of the baseline %human-designed 
SER model. Additionally, Table~\ref{tab:baseline_vs_nas_cnn_only} illustrates the performance of the DARTS-generated model with eight cells ($C=8$), showing a lower performance compared to its counterpart with $C=4$. This decline in performance with an increased number of cells indicates a rise in the model's complexity, leading to overfitting and subsequent accuracy reduction. 
%observation aligns with the notion that models perform better when they exhibit lower complexity~\cite{Sun2023EmotionNAS:Recognition}.

We further examine the results from the CNN branch of Sun et al.'s EmotionNAS model~\cite{Sun2023EmotionNAS:Recognition} to highlight performance enhancements. For a direct and clear comparison of performance, we specifically utilise the `Spectrogram Branch' of EmotionNAS, contrasting it with our `CNN -- DARTS' model. This focused comparison is chosen to ensure a fair evaluation since both models share a similar architecture. 
We see that the performance of our CNN -- DARTS models surpasses the performance of the CNN branch of EmotionNAS by at least 5\%.
It is worth noting that the `whole model' of EmotionNAS has a different architecture, employing a branched structure, while emoDARTS utilises a stacked architecture.

\begin{table}[!ht]
\renewcommand{\arraystretch}{\tablearraystrech}
\caption{Performance comparison between the DARTS generated CNN model (CNN -- DARTS) and a CNN SER model developed without DARTS (CNN -- w/o DARTS) for the IEMOCAP dataset. \\ \textit{The number of parameters is in thousands}}
    \label{tab:baseline_vs_nas_cnn_only}
    \centering
\begin{tabular}{p{0.23\linewidth}rrcc}
\hline 
\textbf{Model}                      & \multicolumn{1}{l}{\textbf{Param.}} & \multicolumn{1}{l}{\textbf{Cell}}       & \textbf{UA (\%)}              & \textbf{WA (\%)}      \\ \hline
CNN -- DARTS                        & \textbf{417K}                   & 4                                         & \textbf{69.36 $\pm$ 3.00}     & $72.55 \pm 3.70$      \\
CNN -- DARTS                        & 428K                            & 8                                         & $62.25 \pm 6.74$              & $63.78 \pm 6.83$      \\
CNN -- w/o DARTS                          & 35K                            & -                                         & $50.04 \pm 2.69$              & $51.01 \pm 2.23$      \\
EmotionNAS [CNN]~\cite{Sun2023EmotionNAS:Recognition}                    & 130K                            & 3                                         & 57.3                          & 63.2                  \\ 
% \small{EmotionNAS [CNN + RNN]}      & 2.37M                             & 3                                         & 69.1                          & 72.1                  \\ 
\hline
\end{tabular}
% \vspace{-15pt}
\renewcommand{\arraystretch}{1}
\end{table}

\subsection{emoDARTS model}

\begin{figure}[!t]
    \centering
    \includegraphics[width=\linewidth]{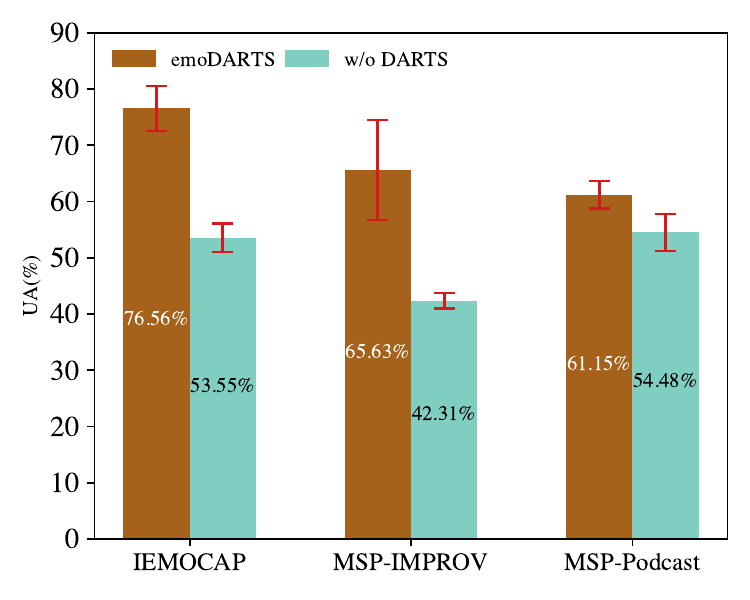}
    \caption{Comparison of UA\% between the datasets the NAS generated (emoDARTS) and CNN+LSTM attention models developed without DARTS (w/o DARTS)}
    \label{fig:ua_comparison}
\end{figure}

We analyse the performance of the CNN-SeqNN model generated by DARTS (emoDARTS) in contrast to the SER models optimised without DARTS (w/o DARTS) %human-designed model (HD) 
and visualise this in Figure~\ref{fig:ua_comparison} and Table~\ref{tab:emodarts_ua_and_wa}.
%as tabulated in Table~\ref{tab:baseline_vs_nas_cnn_lstm}. The table 
The graph shows that the NAS-generated SER model performs better than the 
% human-designed 
baseline SER model developed without DARTS for the three datasets. 

\renewcommand{\arraystretch}{\tablearraystrech}

\begin{table}[!t]
    \centering
    \caption{Unweighted Accuracy (UA\%) of the CNN+LSTM attention model developed without DARTS (w/o DARTS), Unweighted Accuracy (UA\%), and Weighted Accuracy (WA\%) of the emoDARTS model for each Dataset}
    \label{tab:emodarts_ua_and_wa}
    \begin{tabular}{l|c|cc}
        \multirow{2}{*}{\textbf{Dataset}} & \textbf{w/o DARTS} & \multicolumn{2}{c}{\textbf{emoDARTS}}     \\
                                 & UA\%           & \multicolumn{1}{c|}{UA\%} & WA\% \\ \hline
        IEMOCAP                  & 53.55 $\pm$ 2.53             & \multicolumn{1}{l|}{76.56 $\pm$ 4.03}   & 78.03 $\pm$ 3.51   \\
        MSP-IMPROV               & 42.31 $\pm$ 1.34             & \multicolumn{1}{l|}{65.63 $\pm$ 8.85}   & 65.32 $\pm$ 8.73   \\
        MSP-Podcast              & 54.48 $\pm$ 3.25             & \multicolumn{1}{l|}{61.15 $\pm$ 2.41}   & 62.33 $\pm$ 1.80  \\ \hline
    \end{tabular}
\end{table}

\renewcommand{\arraystretch}{1}

We also compare the performance of the SER model generated by our approach with the most related studies, `EmotionNAS' of Sun H.\ et al., and the `CNN\_RNN\_att' system of Wu X.\ et al.\ in Table~\ref{tab:emodarts_vs_lit}. 
\renewcommand{\arraystretch}{\tablearraystrech}
\begin{table}[!t]
\caption{Accuracy of SER models published by related studies compared with our study for the improvised subset in the IEMOCAP dataset.}
    \label{tab:emodarts_vs_lit}
    \centering
    \begin{tabular}{l|c|c}
        \textbf{Study} & \textbf{UA\%} & \textbf{WA\%} \\ \hline
        Sun H.\ et al. 2022 \cite{Sun2023EmotionNAS:Recognition} & 69.1 & 72.1\\
        Wu X.\ et al. 2022 \cite{Wu2022NEURALRECOGNITION} & 56.28 & 68.87\\
        emoDARTS (Our Study) & \textbf{76.56 $\pm$ 4.03} & 78.03 $\pm$ 3.51\\ \hline
    \end{tabular}
\end{table}
\renewcommand{\arraystretch}{1}
%BS: for reproducibility by others, please make sure this "improvised" set can be accessed by others :)
%TR: The IEMOCAP dataset has two main methods: script and impro. Dataset users can directly identify the improv as it is in a separate directory. 
It is visible that the SER model generated by our methodology for the IEMOCAP dataset outperforms the SER models generated by DARTS in the related literature. 

\begin{figure*}
    \centering
    \includegraphics[width=1\linewidth]{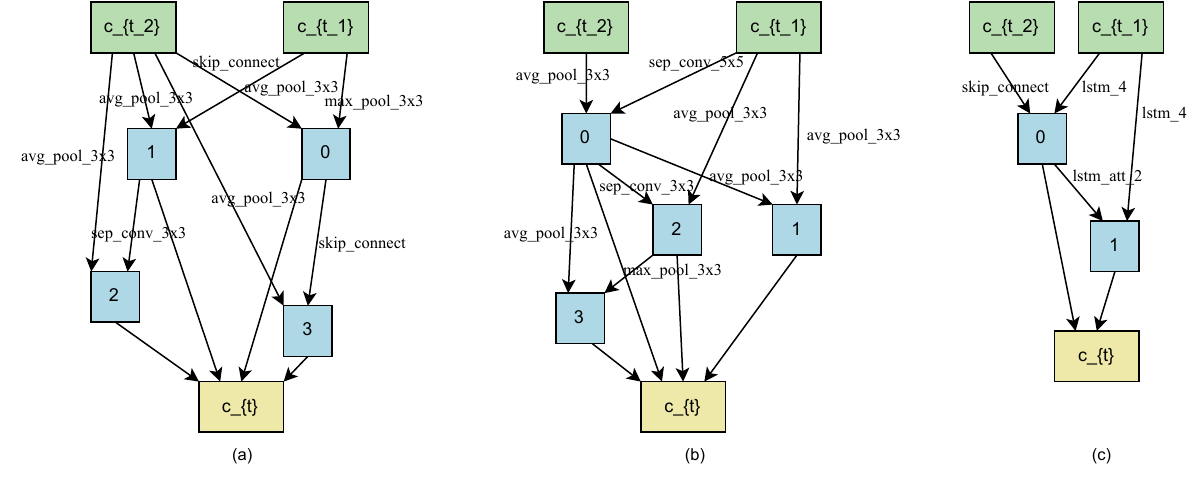}
    \caption{DARTS searched $t^{th}$ cell structure for the CNN Normal Cell (a), CNN Reduction Cell (b), and SeqNN cell (c) for the emoDARTS model.}
    \label{fig:cell_viz}
\end{figure*}

It is further worthwhile to investigate the rationale for increased performance when compared to the results of Wu et al.'s `CNN RNN att'~\cite{Wu2022NEURALRECOGNITION} system. We suggest the improved performance is due to the relaxed candidate operations order rather than the pre-defined layer order. In Wu et al.'s study~\cite{Wu2022NEURALRECOGNITION}, for example, the initial layers are pre-defined to be convolutional layers. The DARTS algorithm must select the best convolutional layer from a pool of just CNN layers. In contrast, our technique allows DARTS to choose among many operations such as convolutions, pooling, and skip connections. Figure~\ref{fig:cell_viz} shows one such use scenario, in which the DARTS searched architecture consists of pooling layers in the initial segments. 

% \HL{The higher accuracy is because of our methodology and not setting the layer order ....}

% Figure~\ref{fig:ua_comparison} also compares the results obtained by the CNN-only model and the CNN+LSTM model. Comparatively, the NAS-generated SER models perform better than hand-engineered SER models. Figure~\ref{fig:cell_viz} shows the searched normal and reduction $t^{th}$ cell (c\_\{t\}) structures for the CNN+LSTM SER model.

Figure~\ref{fig:cell_viz} shows a visualisation of the architecture for each type of cell (normal and reduction cell of the CNN component, and cell in the SeqNN component) searched by DARTS for the emoDARTS model. It is visible that  DARTS has selected three LSTM based operations for the SeqNN component and only one of them contains attention. This shows that jointly optimising the emoDARTS model has enabled the DARTS framework to choose optimum operations rather than blindly choosing layers with `attention' for all the operations. 

\subsection{Restricting the search scope}
We study the impact on the performance of the searched model by restricting the search scope for the SeqNN component. We divide the search scope into five segments namely `LSTM Only', `LSTM-Att. Only', `RNN Only', and `RNN-Att. Only'. Table~\ref{tab:restricting_seqnn_search_scope} shows the DARTS operations allowed as the candidate operations in the SeqNN component during the search phase. 

\renewcommand{\arraystretch}{\tablearraystrech}
\begin{table}[!ht]
    \caption{DARTS operations allowed as the candidate operations in the SeqNN component during the search phase}
    \label{tab:restricting_seqnn_search_scope}
    \centering
    \begin{tabular}{p{0.235\linewidth}|p{0.6\linewidth}}
        \textbf{Scope} & \textbf{Candidate Operations} \\ \hline
        emoDARTS & lstm\_1, lstm\_2, lstm\_3, lstm\_4, lstm\_att\_1, lstm\_att\_2, rnn\_1, rnn\_2, rnn\_3, rnn\_4, rnn\_att\_1, rnn\_att\_2 \\ \hline
        LSTM Only & lstm\_1, lstm\_2, lstm\_3, lstm\_4 \\ \hline
        LSTM-Att. Only & lstm\_att\_1, lstm\_att\_2 \\ \hline
        RNN Only & rnn\_1, rnn\_2, rnn\_3, rnn\_4 \\ \hline
        RNN-Att. Only & rnn\_att\_1, rnn\_att\_2 \\ \hline
    \end{tabular}

\end{table}
\renewcommand{\arraystretch}{1}

Table~\ref{tab:restricting_candidate_genome_ops} shows the performance and number of parameters of the searched model when the candidate search operations are restricted. Here, we study the effect on the performance of the 
searched architecture when the search algorithm was only given a restricted set of operations. For example, the `LSTM Only' study only allowed to use operations from lstm\_1, lstm\_2, lstm\_3, and lstm\_4. We try to identify the most important types of genome operations that we can use in the search algorithm. This approach allows to use of only the important operations in the search scope and optimises the memory utilisation in the search phase. 

\renewcommand{\arraystretch}{\tablearraystrech}
\begin{table*}[!t]
\centering
\caption{Performance (UA\%, WA\%) and number of parameters (Param.) of each generated model when the candidate search operations are restricted}
\label{tab:restricting_candidate_genome_ops}
\setlength{\tabcolsep}{4.5pt}
% \newcolumntype{|}{!{\vrule width 1pt}}
\begin{tabular}{lccr|ccr|ccr}
\multirow{2}{*}{}   & \multicolumn{3}{c|}{\textbf{IEMOCAP}}                                                 & \multicolumn{3}{c|}{\textbf{MSP-Improv}}     & \multicolumn{3}{c}{\textbf{MSP-Podcast}}                                             \\
           \textbf{Genome Ops.}         & \multicolumn{1}{c}{\textbf{UA\%}} & \multicolumn{1}{c}{\textbf{WA\%}} & \multicolumn{1}{c|}{\textbf{Param.}} & \multicolumn{1}{c}{\textbf{UA\%}} & \multicolumn{1}{c}{\textbf{WA\%}} & \multicolumn{1}{c|}{\textbf{Param.}} & \multicolumn{1}{c}{\textbf{UA\%}} & \multicolumn{1}{c}{\textbf{WA\%}} & \multicolumn{1}{c}{\textbf{Param.}} \\ \hline
emoDARTS      & \textbf{76.55 $\pm$ 4.03} & 78.03 $\pm$ 3.51                               & 1\,014\,556                                       & \textbf{65.63 $\pm$ 8.85}   & 65.32 $\pm$ 8.73                          & 1\,008\,812    & \textbf{61.15 $\pm$ 2.41}  & 62.33 $\pm$ 1.79                             & 1\,835\,604                                  \\
LSTM Only           & 67.42 $\pm$ 5.76 & 67.78 $\pm$ 5.97                               & 2\,939\,372                                       & 32.35 $\pm$ 1.93 & 31.85 $\pm$ 0.98                                & 2\,931\,748    & 53.12 $\pm$ 5.82     & 54.48 $\pm$ 5.43                          & 2\,943\,852                                  \\
LSTM-Att. Only & 68.12 $\pm$ 7.12     & 68.59 $\pm$ 6.62                           & 7\,284\,412                                       & 49.74 $\pm$ 10.08 & 49.8 $\pm$ 10.15                                & 6\,474\,676     & 53.11 $\pm$ 5.83   & 55.07 $\pm$ 5.14                             & 7\,284\,932                                 \\
RNN Only             & 70.42 $\pm$ 8.32  & 72.07 $\pm$ 8.23                               & 356\,492                                        & 63.59 $\pm$ 7.88 & 63.49 $\pm$ 7.95                               & 1\,016\,284     & 59.16 $\pm$ 6.67      & 60.72 $\pm$ 7.65                          & 706\,068                                 \\
RNN-Att. Only & 36.14 $\pm$ 10.97  & 48.63 $\pm$ 6.56                              & 29\,324                                         & 57.71 $\pm$ 12.61 & 57.04 $\pm$ 12.14                                 & 22\,932      & 59.29 $\pm$ 6.38 & 60.25 $\pm$ 6.09                               & 2\,844\,420                                  \\ \hline
\end{tabular}
\end{table*}
\renewcommand{\arraystretch}{1}

\begin{figure*}[!ht]
    \centering
    \includegraphics[width=1\textwidth]{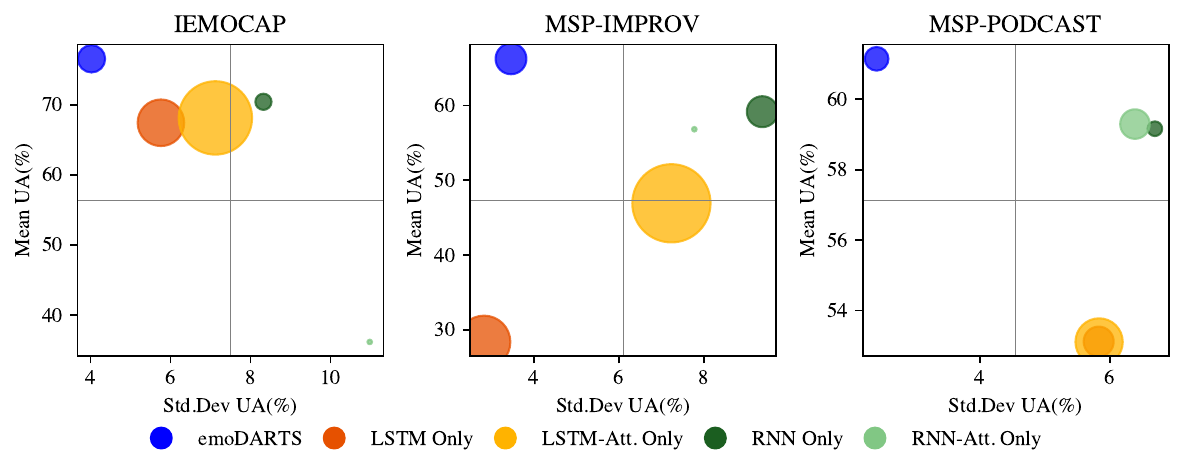}
    \caption{Visualisation of results for the studies restricting the search space for the three datasets: IEMOCAP, MSP-IMPROV, and MSP-Podcast. The vertical axis is the mean UA\% and the horizontal axis is the standard deviation of UA\%. The size of the marker depicts the size (number of parameters) of the generated model. The best performing model can be found at the top-left most position of the figure which has the highest mean UA\% and lowest standard deviation of UA\%.}
    \label{fig:restricting_candidate_genome_ops}
\end{figure*}

Comparing the trials `emoDARTS' and `LSTM Only' in the IEMOCAP dataset, we can observe that even though the number of parameters has tripled in the `LSTM Only' scenario, the performance (UA\%) has not increased. This indicates that increasing the number of parameters just by increasing the complexity of the model does not tend to give better performance, but the model components should be compatible with each other. 

\begin{figure}[!t]
    \centering
    \includegraphics[width=1\linewidth]{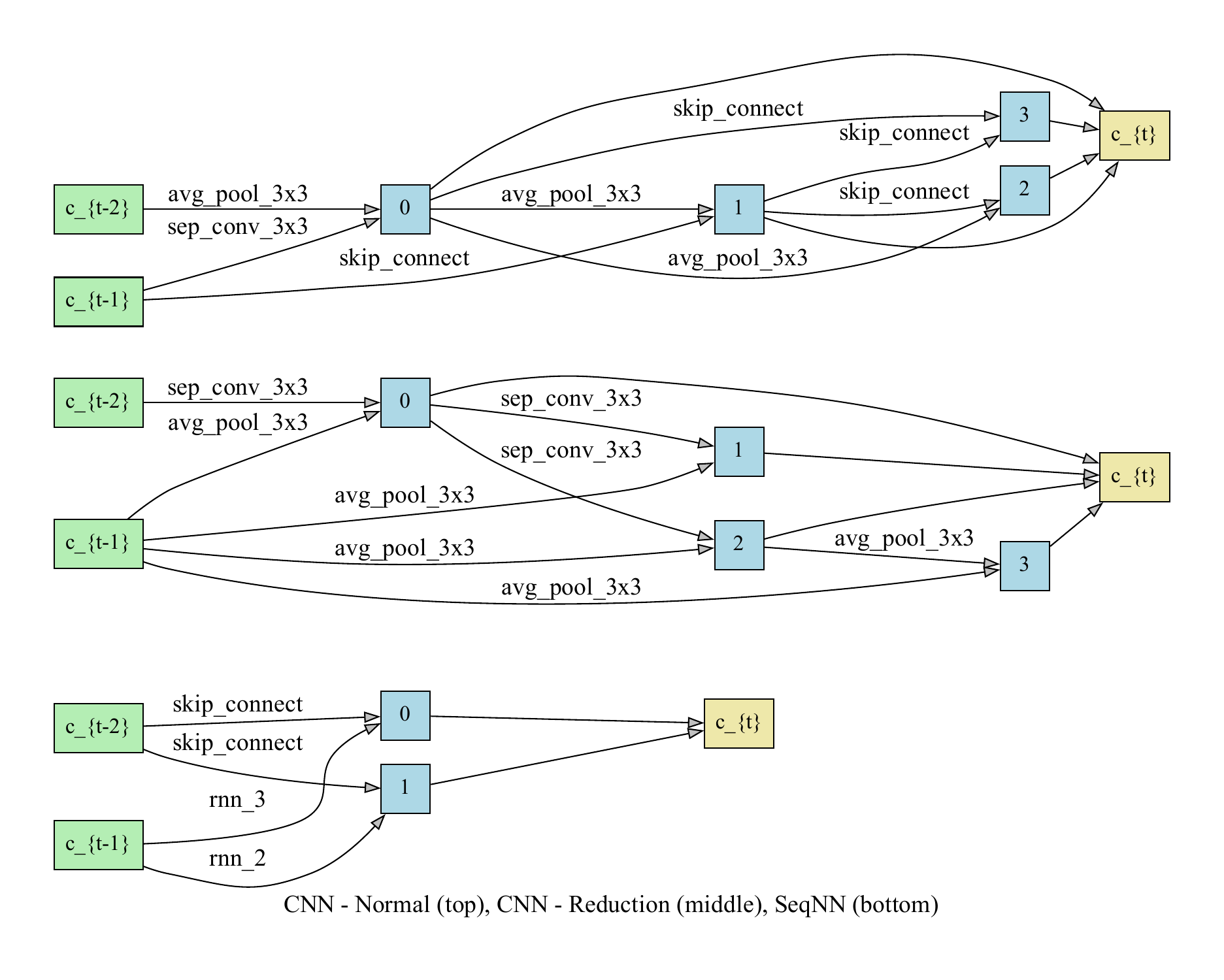}
    \caption{DARTS searched cell structure for CNN and SeqNN cells when the SeqNN search space has only RNN operations (RNN Only) for the IEMOCAP dataset}
    \label{fig:rnn_only_genome}
\end{figure}

Notably, models using 'RNN Only' genomic operations achieve the second-highest accuracy despite having much fewer trainable parameters. Figure~\ref{fig:rnn_only_genome} depicts the cell architecture, which consists mostly of pooling layers and skip connection operations that do not have any training parameters and hence do not contribute to the total number of trainable parameters.

We provide in Table~\ref{tab:restricting_candidate_genome_ops} results as well as a scatter plot for better visualisation in Figure~\ref{fig:restricting_candidate_genome_ops}, where the mean UA\% in the vertical axis, standard deviation of UA\% in the horizontal axis and size of the markers indicates the number of parameters. The polts indicate which model gives better performance in terms of the mean accuracy and its standard deviation. The best performing model can be found in the top left corner of the plot where the highest UA\% and lowest Standard deviation of UA\% are present. According to the figure, the best performing model for all three datasets is given by `emoDARTS' where all the candidate operations are available in the search scope. 

\section{Discussion}
Throughout this study, we encountered various challenges. In this section, we report the key challenges and our strategies for overcoming them. The three primary challenges we faced were:
\begin{enumerate}
    \item Optimising the GPU Memory utilisation % High GPU Memory Usage in the Search phase
    \item Converging to a local minima
    \item High Standard Deviation of the results 
\end{enumerate}

\subsection{Optimising the GPU Memory utilisation}
% As the DARTS algorithm models the Search problem as a network graph, it creates multiple edges between each node. The number of edges between each node equals the candidate operations defined. 
% An operation could be a simple CNN layer, Pooling layer, RNN layer or a complex module such as an LSTM-attention module. 
% During the search phase, a super-neural network is created. This means the super-neural network has multiple instances of neural network layers or modules. 

The DARTS algorithm conceptualises the search problem as a network graph, establishing multiple edges between each node. The quantity of edges corresponds to the defined candidate operations. These operations encompass various possibilities, ranging from simple CNN, pooling, and RNN layers to intricate modules like an LSTM-attention module. In the search phase, a super-neural network is constructed, resulting in multiple instances of neural network layers or modules within this overarching structure.

\begin{figure}[!t]
    \centering
    \includegraphics[width=1\linewidth]{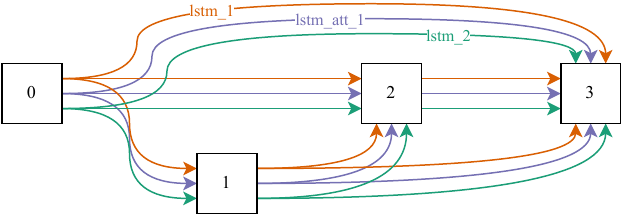}
    \caption{Example graph of a DARTS cell which has four nodes and candidate operations are ``lstm\_1'', ``lstm\_2'', and ``lstm\_att\_1''. The same edge colour denotes the same type of operation.}
    \label{fig:supernet_example_lstm}
\end{figure}

Figure~\ref{fig:supernet_example_lstm} shows an example of a graph inside a DARTS cell that has four nodes and the candidate operations are ``lstm\_1'', ``lstm\_2'', and ``lstm\_att\_1'', where ``lstm\_1'' is a single layer LSTM component, ``lstm\_2'' is a double layer LSTM component, and ``lstm\_att\_1'' is an attention induced single layer LSTM component.
According to the example, a single DARTS cell should initiate 6$\times$lstm\_1 layers, 6$\times$lstm\_2 layers, and 6$\times$lstm\_att\_1 layers. If the search configuration has $4$ cells, we have to initialise $4$ instances of cells where all the weight and bias parameters have to be initialised in the computing device. This will increase the GPU memory utilisation. 

Providing a higher number of nodes in a cell, a higher number of cells, and expanding the array of candidate operations will increase the amount of GPU memory utilisation and eventually will exhaust the GPU memory capacity failing the search operation. 

To optimise the GPU memory utilisation, we recommend conducting an assessment to determine the set of possible search operations and hyperparameters such as the number of layers, cells, and nodes inside the cell considering the GPU resources available.

On the other hand, based on the results of Table~\ref{tab:restricting_candidate_genome_ops}, we should not be restricted only to a single type of network but rather should consist of a variety of network architectures. % \RR{Can you give specific example. Readers might not understand what you are referring them in Table 7}  
We selected the set of candidate operations indicated in Table~\ref{tab:restricting_seqnn_search_scope}
%BS: please avoid hard references ;)
under 'emoDARTS' based on GPU resource availability and on the premise that all sorts of candidate operations should be available in the search space.

\subsection{Converging to a local minima}
\label{sec:converging_to_local}
\begin{figure}[!t]
    \centering
    \includegraphics[width=1\linewidth]{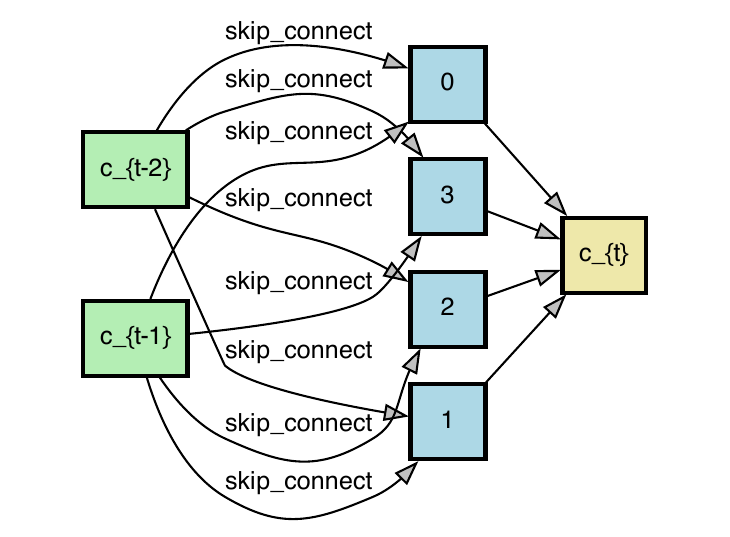}
    \caption{DARTS genome of the SeqNN module where the search algorithm selected the identity operation as all the operations.}
    \label{fig:local_minima_rnn}
\end{figure}

Throughout the course of our experiments, we attempted various configurations for the number of cells and nodes. We observed that the SeqNN module converges to a local minimum when the number of cells and number of nodes is greater than $3$. The output of the searched genome for the SeqNN module contained all ``skip\_connect'' which indicates identity operations are used instead of any RNN or LSTM operations. Figure~\ref{fig:local_minima_rnn} shows one such instance DARTS SeqNN genome.

We were able to address the challenge by reducing the complexity of the candidate search graph by reducing the number of cells and the number of nodes inside a cell. More research is, however, needed to manage a more complex search network.

\subsection{High Standard Deviation in the results}
An important observation derived from our results is the high standard deviation. This can be attributed to the dataset-splitting method we employed. Specifically, we adopt speaker-independent dataset splitting, where the training and validation sets are segregated based on the speaker. In this configuration, any audio utterance from a particular speaker in the validation set remains unseen by the model during training. Consequently, the DARTS-optimised model is not trained to handle the data distribution of the validation set. To tackle this challenge, potential solutions include dataset poisoning and enhancing the generalisation capabilities of the SER model by incorporating dropout layers.

% After evaluating the results, we see that the standard deviation of the performance results we received is higher. Dataset splitting explains this phenomenon. We consider speaker-independent dataset splitting, which separates the training and validation sets based on the speaker. Any audio utterance of one speaker in the validation set is not exposed to the model during training. As a result, the DARTS-optimised model is not trained to handle the validation set's data distribution. Possible solutions to this problem include 1. Dataset poisoning. 2. Improving the SER model's generalisation capabilities by incorporating dropout layers. 

\section{Conclusions}
%BS: TEMPUS: ALL should be in PRESENT TENS, but the Conclusion in PAST TENSE. Here, it was mixed or almost inverted - I changed throughout :)

% This research aimed to determine the viability of utilising the DARTS algorithm to search a neural architecture for a joint CNN, SeqNN SER model. We use DARTS to optimise the whole architecture for an optimum-performing SER model jointly. The performance results shows that the 'emoDARTS` model performs better than the hand engineered models with CNN and LSTM. 

% \TJ{Change CNN+LSTM to CNN+RNN and include attention mechnism for RNN} 
% This research aimed to study the feasibility of utilising the DARTS algorithm to search a neural architecture for a CNN+LSTM SER model.
% \RR{you need to tell here why you just optimised CNN using DARTS not both.} 
% \RR{You also need to talk about the number of parameters as a metric - this has no focus whatsoever. }
% The results indicate that DARTS optimises the neural architecture to outperform hand-engineered models. Furthermore, it reveals that as complexity increases, performance decreases, leading to the SER model performing better when it is simpler
% \RR{I don't understand what you are trying to say here - this needs to be expanded}. 
% The future work of this study is to optimise the SER model architecture using attention networks and using DARTS to design the combined CNN+LSTM model with optimum performance and minimum complexity. 
% \RR{you need to rewrite the conclusion - it looks half-baked}

%BS: conclusion in past tense - I changed.
In conclusion, this paper introduced an innovative approach to enhancing speech emotion recognition (SER) using differentiable architecture search by DARTS. Our primary focus was on tailoring DARTS for a joint configuration of a  Convolutional Neural Network (CNN) and a Sequential Neural Network, deviating from previous studies by allowing DARTS to autonomously determine the optimal layer order for the CNN within the DARTS cell without imposing constraints.

A comprehensive evaluation was conducted, comparing our proposed method with
baseline models developed without DARTS
% human-designed network alternatives 
and various genome operations, including LSTM only, LSTM with attention only, RNN only, and RNN with attention only. The detailed assessments consistently demonstrate the superior performance of our proposed method. Contrasting with existing studies further validates the effectiveness and superiority of our approach, considering parameter size and accuracy as essential dimensions for comparison.

Notably, our study extends beyond the confines of the commonly used IEMOCAP dataset, incorporating two additional datasets, MSP-IMPROV and MSP-Podcast. This extension showcases the superior performance of our proposed method across diverse datasets, affirming its generalisation capability.

Furthermore, we shared valuable insights gained from our experiences, addressing challenges related to GPU exhaustion and converging to local minima. These insights serve as practical guidance for researchers, helping them navigate potential pitfalls and optimise the application of DARTS in SER.

%BS: added an outlook
Future efforts will need to deal with neural architecture for further modern architectures such as transformers and translating the made findings beyond the targeted field of application.

%BS: Please check the references - they look quite mixed such as 31 written all in capitals...

\bibliographystyle{IEEEtran}

\bibliography{references}

% Generated by IEEEtran.bst, version: 1.13 (2008/09/30)
\begin{thebibliography}{10}
\providecommand{\url}[1]{#1}
\csname url@samestyle\endcsname
\providecommand{\newblock}{\relax}
\providecommand{\bibinfo}[2]{#2}
\providecommand{\BIBentrySTDinterwordspacing}{\spaceskip=0pt\relax}
\providecommand{\BIBentryALTinterwordstretchfactor}{4}
\providecommand{\BIBentryALTinterwordspacing}{\spaceskip=\fontdimen2\font plus
\BIBentryALTinterwordstretchfactor\fontdimen3\font minus \fontdimen4\font\relax}
\providecommand{\BIBforeignlanguage}[2]{{%
\expandafter\ifx\csname l@#1\endcsname\relax
\typeout{** WARNING: IEEEtran.bst: No hyphenation pattern has been}%
\typeout{** loaded for the language `#1'. Using the pattern for}%
\typeout{** the default language instead.}%
\else
\language=\csname l@#1\endcsname
\fi
#2}}
\providecommand{\BIBdecl}{\relax}
\BIBdecl

\bibitem{Zhao2019SpeechNetworks}
J.~Zhao, X.~Mao, and L.~Chen, ``{Speech emotion recognition using deep 1D {\&} 2D CNN LSTM networks},'' \emph{Biomedical Signal Processing and Control}, vol.~47, pp. 312--323, 1 2019.

\bibitem{Jalal2020EmpiricalRecognition}
M.~A. Jalal, R.~Milner, and T.~Hain, ``{Empirical interpretation of speech emotion perception with attention based model for speech emotion recognition},'' \emph{Proceedings of the Annual Conference of the International Speech Communication Association, INTERSPEECH}, vol. 2020-October, pp. 4113--4117, 2020.

\bibitem{Lieskovska2021AMechanism}
E.~Lieskovsk{\'{a}}, M.~Jakubec, R.~Jarina, M.~Chmul{\'{i}}k, Y.-F. Liao, P.~Bours, and C.~Kwan, ``{A Review on Speech Emotion Recognition Using Deep Learning and Attention Mechanism},'' \emph{Electronics 2021, Vol. 10, Page 1163}, vol.~10, no.~10, p. 1163, 5 2021.

\bibitem{Latif2022MultitaskRecognition}
\BIBentryALTinterwordspacing
S.~Latif, R.~Rana, S.~Khalifa, R.~Jurdak, and B.~W. Schuller, ``{Multitask Learning From Augmented Auxiliary Data for Improving Speech Emotion Recognition},'' \emph{IEEE Transactions on Affective Computing}, pp. 1--13, 7 2022. [Online]. Available: \url{https://ieeexplore.ieee.org/document/9947296/}
\BIBentrySTDinterwordspacing

\bibitem{Han2018TowardsRecognition}
W.~Han, H.~Ruan, X.~Chen, Z.~Wang, H.~Li, and B.~Schuller, ``{Towards temporal modelling of categorical speech emotion recognition},'' \emph{Proceedings of the Annual Conference of the International Speech Communication Association, INTERSPEECH}, vol. 2018-September, pp. 932--936, 2018.

\bibitem{Haque2018ImageAttention}
\BIBentryALTinterwordspacing
K.~N. Haque, M.~A. Yousuf, and R.~Rana, ``{Image denoising and restoration with CNN-LSTM Encoder Decoder with Direct Attention},'' \emph{arXiv Prepr.}, pp. 1--12, 1 2018. [Online]. Available: \url{https://arxiv.org/abs/1801.05141v1}
\BIBentrySTDinterwordspacing

\bibitem{Li2019ImprovedLearning}
Y.~Li, T.~Zhao, and T.~Kawahara, ``{Improved end-to-end speech emotion recognition using self attention mechanism and multitask learning},'' \emph{Proceedings of the Annual Conference of the International Speech Communication Association, INTERSPEECH}, vol. 2019-September, pp. 2803--2807, 2019.

\bibitem{Latif2022SelfRecognition}
S.~Latif, R.~Rana, S.~Khalifa, R.~Jurdak, and B.~W. Schuller, ``{Self Supervised Adversarial Domain Adaptation for Cross-Corpus and Cross-Language Speech Emotion Recognition},'' \emph{IEEE Transactions on Affective Computing}, 2022.

\bibitem{Sun2023EmotionNAS:Recognition}
H.~Sun, Z.~Lian, B.~Liu, Y.~Li, L.~Sun, C.~Cai, J.~Tao, M.~Wang, and Y.~Cheng, ``{EmotionNAS: Two-stream Neural Architecture Search for Speech Emotion Recognition},'' \emph{Proceedings of the Annual Conference of the International Speech Communication Association, INTERSPEECH}, vol. 2023-August, pp. 3597--3601, 2023.

\bibitem{Wu2022NEURALRECOGNITION}
X.~Wu, S.~Hu, Z.~Wu, X.~Liu, and H.~Meng, ``{Neural Architecture Search for Speech Emotion Recognition},'' \emph{ICASSP, IEEE International Conference on Acoustics, Speech and Signal Processing - Proceedings}, vol. 2022-May, pp. 6902--6906, 2022.

\bibitem{Xu2020PC-DARTS:Search}
Y.~Xu, L.~Xie, X.~Zhang, X.~Chen, G.-J. Qi, Q.~Tian, and H.~Xiong, ``{PC-DARTS: Partial Channel Connections for Memory-Efficient Architecture Search},'' in \emph{International Conference on Learning Representations}, 7 2020.

\bibitem{Biju2019ElectricDARTS}
G.~M. Biju, G.~N. Pillai, and J.~Seshadrinath, ``{Electric load demand forecasting with RNN cell generated by DARTS},'' \emph{IEEE Region 10 Annual International Conference, Proceedings/TENCON}, vol. 2019-October, pp. 2111--2116, 10 2019.

\bibitem{Maulik2020RecurrentEmulation}
R.~Maulik, R.~Egele, C.~Polytechnique, B.~Lusch, and P.~Balaprakash, ``{Recurrent Neural Network Architecture Search for Geophysical Emulation},'' \emph{Proceedings of the International Conference for High Performance Computing, Networking, Storage and Analysis}, 2020.

\bibitem{Zheng2015AnNetworks}
W.~Q. Zheng, J.~S. Yu, and Y.~X. Zou, ``{An experimental study of speech emotion recognition based on deep convolutional neural networks},'' \emph{2015 International Conference on Affective Computing and Intelligent Interaction, ACII 2015}, pp. 827--831, 12 2015.

\bibitem{Trigeorgis16-AFE}
G.~Trigeorgis, F.~Ringeval, R.~Brueckner, E.~Marchi, M.~A. Nicolaou, B.~Schuller, and S.~Zafeiriou, ``{Adieu features? End-to-end speech emotion recognition using a deep convolutional recurrent network},'' \emph{ICASSP, IEEE International Conference on Acoustics, Speech and Signal Processing - Proceedings}, pp. 5200--5204, 5 2016.

\bibitem{Zhang18-ELF}
Z.~Zhang, J.~Han, K.~Qian, and B.~Schuller, ``{Evolving Learning for Analysing Mood-Related Infant Vocalisation},'' \emph{Proceedings INTERSPEECH 2018, 19. Annual Conference of the International Speech Communication Association}, pp. 142--146, 2018.

\bibitem{Zoph2016NeuralLearning}
B.~Zoph and Q.~V. Le, ``{Neural Architecture Search with Reinforcement Learning},'' \emph{5th International Conference on Learning Representations, ICLR 2017 - Conference Track Proceedings}, 11 2016.

\bibitem{Liberis2020NAS:Microcontrollers}
E.~Liberis, L.~Dudziak, and N.~D. Lane, ``{{$\mu$}NAS: Constrained Neural Architecture Search for Microcontrollers},'' \emph{Proceedings of the 1st Workshop on Machine Learning and Systems, EuroMLSys 2021}, pp. 70--79, 10 2020.

\bibitem{Gong2019MixedLearning}
C.~Gong, Z.~Jiang, D.~Wang, Y.~Lin, Q.~Liu, and D.~Z. Pan, ``{Mixed precision neural architecture search for energy efficient deep learning},'' \emph{IEEE/ACM International Conference on Computer-Aided Design, Digest of Technical Papers, ICCAD}, vol. 2019-November, 11 2019.

\bibitem{Liu2018DARTS:SEARCH}
\BIBentryALTinterwordspacing
H.~Liu, K.~Simonyan, and Y.~Yang, ``{DARTS: Differentiable Architecture Search},'' \emph{7th International Conference on Learning Representations, ICLR 2019}, 6 2018. [Online]. Available: \url{https://arxiv.org/abs/1806.09055v2}
\BIBentrySTDinterwordspacing

\bibitem{Liu2018ProgressiveSearch}
C.~Liu, B.~Zoph, M.~Neumann, J.~Shlens, W.~Hua, L.-J. Li, L.~Fei-Fei, A.~Yuille, J.~Huang, and K.~Murphy, ``{Progressive Neural Architecture Search},'' in \emph{Proceedings of the European Conference on Computer Vision (ECCV)}, 2018, pp. 19--34.

\bibitem{Liang2019DARTS+:Stopping}
H.~Liang, S.~Zhang, J.~Sun, X.~He, W.~Huang, K.~Zhuang, and Z.~Li, ``{DARTS+: Improved Differentiable Architecture Search with Early Stopping},'' \emph{arXiv preprint}, 9 2019.

\bibitem{Wan2020FBNetV2:Dimensions}
A.~Wan, X.~Dai, P.~Zhang, Z.~He, Y.~Tian, S.~Xie, B.~Wu, M.~Yu, T.~Xu, K.~Chen, P.~Vajda, and J.~E. Gonzalez, ``{FBNetV2: Differentiable Neural Architecture Search for Spatial and Channel Dimensions},'' in \emph{Proceedings of the IEEE/CVF Conference on Computer Vision and Pattern Recognition (CVPR)}, 2020, pp. 12\,965--12\,974.

\bibitem{Chen2020StabilizingRegularization}
X.~Chen and C.-J. Hsieh, ``{Stabilizing Differentiable Architecture Search via Perturbation-based Regularization},'' \emph{Proceedings of the 37th International Conference on Machine Learning}, 2020.

\bibitem{Yu2023CyclicSearch}
H.~Yu, H.~Peng, Y.~Huang, J.~Fu, H.~Du, L.~Wang, and H.~Ling, ``{Cyclic Differentiable Architecture Search},'' \emph{IEEE Transactions on Pattern Analysis and Machine Intelligence}, vol.~45, no.~1, pp. 211--228, 1 2023.

\bibitem{Rajapakshe2023EnhancingSearch}
\BIBentryALTinterwordspacing
T.~Rajapakshe, R.~Rana, S.~Khalifa, B.~Sisman, and B.~W. Schuller, ``{Enhancing Speech Emotion Recognition Through Differentiable Architecture Search},'' \emph{arXiv preprint}, 5 2023. [Online]. Available: \url{https://arxiv.org/abs/2305.14402v3}
\BIBentrySTDinterwordspacing

\bibitem{Zhang2018AttentionRecognition}
Y.~Zhang, J.~Du, Z.~Wang, J.~Zhang, and Y.~Tu, ``{Attention Based Fully Convolutional Network for Speech Emotion Recognition},'' \emph{2018 Asia-Pacific Signal and Information Processing Association Annual Summit and Conference, APSIPA ASC 2018 - Proceedings}, pp. 1771--1775, 7 2018.

\bibitem{Chen20183-DRecognition}
M.~Chen, X.~He, J.~Yang, and H.~Zhang, ``{3-D Convolutional Recurrent Neural Networks with Attention Model for Speech Emotion Recognition},'' \emph{IEEE Signal Processing Letters}, vol.~25, no.~10, pp. 1440--1444, 10 2018.

\bibitem{Zou2022SPEECHINFORMATION}
H.~Zou, Y.~Si, C.~Chen, D.~Rajan, and E.~S. Chng, ``{Speech Emotion Recognition with Co-Attention Based Multi-Level Acoustic Information},'' \emph{ICASSP, IEEE International Conference on Acoustics, Speech and Signal Processing - Proceedings}, vol. 2022-May, pp. 7367--7371, 2022.

\bibitem{Liu2023SpeechLearning}
Z.~T. Liu, M.~T. Han, B.~H. Wu, and A.~Rehman, ``{Speech emotion recognition based on convolutional neural network with attention-based bidirectional long short-term memory network and multi-task learning},'' \emph{Applied Acoustics}, vol. 202, p. 109178, 1 2023.

\bibitem{Busso2008IEMOCAP:Database}
C.~Busso, M.~Bulut, C.-C. Lee, A.~Kazemzadeh, E.~Mower, S.~Kim, J.~N. Chang, S.~Lee, and S.~S. Narayanan, ``\BIBforeignlanguage{en}{{IEMOCAP: interactive emotional dyadic motion capture database}},'' \emph{\BIBforeignlanguage{en}{Language Resources and Evaluation}}, vol.~42, no.~4, p. 335, 2008.

\bibitem{Busso2017MSP-IMPROV:Perception}
C.~Busso, S.~Parthasarathy, A.~Burmania, M.~AbdelWahab, N.~Sadoughi, and E.~M. Provost, ``{MSP-IMPROV: An Acted Corpus of Dyadic Interactions to Study Emotion Perception},'' \emph{IEEE Transactions on Affective Computing}, vol.~8, no.~1, pp. 67--80, 2017.

\bibitem{Lotfian2019BuildingRecordings}
R.~Lotfian and C.~Busso, ``{Building Naturalistic Emotionally Balanced Speech Corpus by Retrieving Emotional Speech from Existing Podcast Recordings},'' \emph{IEEE Transactions on Affective Computing}, vol.~10, no.~4, pp. 471--483, 10 2019.

\bibitem{Davis1980ComparisonSentences}
S.~Davis and P.~Mermelstein, ``{Comparison of parametric representations for monosyllabic word recognition in continuously spoken sentences},'' \emph{IEEE transactions on acoustics, speech, and signal processing}, vol.~28, no.~4, pp. 357--366, 1980.

\bibitem{Latif2019DirectSpeech}
S.~Latif, R.~Rana, S.~Khalifa, R.~Jurdak, and J.~Epps, ``{Direct Modelling of Speech Emotion from Raw Speech},'' in \emph{Proceedings of the Annual Conference of the International Speech Communication Association, INTERSPEECH}, 2019, pp. 3920--3924.

\bibitem{McFee2015Librosa:Python}
B.~McFee, C.~Raffel, D.~Liang, D.~P.~W. Ellis, M.~McVicar, E.~Battenberg, and O.~Nieto, ``{librosa: Audio and music signal analysis in python},'' in \emph{Proceedings of the 14th python in science conference}, vol.~8, 2015.

\bibitem{Paszke2019PyTorch:Library}
A.~Paszke, S.~Gross, F.~Massa, A.~Lerer, J.~Bradbury~Google, G.~Chanan, T.~Killeen, Z.~Lin, N.~Gimelshein, L.~Antiga, A.~Desmaison, A.~K. Xamla, E.~Yang, Z.~Devito, M.~Raison~Nabla, A.~Tejani, S.~Chilamkurthy, Q.~Ai, B.~Steiner, L.~F. Facebook, J.~B. Facebook, and S.~Chintala, ``{PyTorch: An Imperative Style, High-Performance Deep Learning Library},'' \emph{Advances in Neural Information Processing Systems}, pp. 8024--8035, 2019.

\end{thebibliography}

\end{document}